\documentstyle[aps,multicol,epsf]{revtex}

\newcommand{\be}{\begin{equation}}
\newcommand{\ee}{\end{equation}}

\begin{document}
\draft
\widetext

\title{Topological interpretation of subharmonic mode locking
 in coupled oscillators with inertia}

\author{M.Y. Choi}
\address
{Department of Physics, Seoul National University, Seoul 151-742, Korea}

\author{D.J. Thouless}
\address{Department of Physics, University of Washington, Seattle, WA 98195}

\maketitle

\begin{abstract}
A topological argument is constructed and applied
to explain subharmonic mode locking in 
a system of coupled oscillators with inertia.
Via a series of transformations, the system is shown to be
described by a classical $XY$ model with periodic bond angles, 
which is in turn mapped onto a tight-binding particle in a periodic
gauge field.  It is then revealed that subharmonic quantization 
of the average phase velocity 
follows as a manifestation of topological invariance. 
Ubiquity of multistability and associated hysteresis are also pointed out.
\end{abstract}


\begin{multicols}{2}
\narrowtext

In a nonlinear oscillator system driven periodically, 
the competition between the 
natural frequency and the driving frequency in general leads to 
either an almost periodic motion or a periodic one, depending on
the parameter range~\cite{modelock}.
The latter, called mode locking, is 
characterized by the quantization at rational values of the 
average phase velocity.
In particular subharmonic mode locking, appearing in the presence of 
the inertia term, results in the devil's staircase structure. 
One of the well-known examples is the Josephson junction 
driven by combined direct and alternating currents, 
with the capacitance playing the role of inertia~\cite{JJ}. 
Governed by the same equation of motion as a driven pendulum, 
it displays dc voltage plateaus in the current-voltage characteristics,
known as Shapiro steps~\cite{SS}. 
Similar voltage quantization has also been 
observed in arrays of Josephson junctions, 
yielding integer giant Shapiro steps~\cite{GSS,FGSS} and subharmonic steps~\cite{sub} 
according to the absence/presence of capacitive terms.
Unfortunately, in spite of the deceptively simple equation of motion, 
even the {\em single}-junction problem has resisted complete analytical solutions,
especially, in the presence of the capacitive term,
except for the results mainly based on the circle map~\cite{Arnold} 
and on the approximate analysis by means of
expansion and averaging~\cite{Minorsky,hys}. 
Accordingly, such mode locking phenomena in {\em arrays} 
have been demonstrated mostly by numerical simulations. 
On the other hand, the topological argument, 
proposed for the system without the capacitive term~\cite{Choi},
reveals topological invariance of the system as the physical origin 
of quantization~\cite{Thouless}. 
As in the case of the quantum Hall effect~\cite{qhe},
the topological argument does not provide 
quantitative information, e.g., on the locking structure. 
Nevertheless it not only clarifies the nature of quantization but also 
provides a link between dynamics and statics 
by interpreting (dynamical) mode locking in terms of
(static) topological invariance. 

In this work, we construct a topological argument
for the system with inertia, 
and apply the idea to Josephson-junction arrays 
or systems of coupled oscillators, 
with attention to the resulting subharmonic locking.
For this purpose, we consider an appropriate canonical transformation
of the dynamic equations of motion and the corresponding Fokker-Planck
equation, the stationary solution of which gives
the effective Hamiltonian in the form
of a classical $XY$ model with periodic bond angles. 
Via mapping onto a tight-binding particle in a periodic gauge field,
subharmonic quantization of the average (dc) phase velocity is revealed as
a manifestation of the topological invariance. 
Also suggested is ubiquity of multistability,
providing a natural explanation of the observed hysteresis due to the inertia.

We begin with the set of equations of motion for
a system of $N$ coupled oscillators 
\be \label{eom}
\sum_{j=1}^N [M_{ij} \ddot{\phi}_j + \gamma M_{ij} \dot{\phi}_j 
    + J_{ij} \sin (\phi_i -\phi_j ) ]
  = I_i ,
\ee
where $\phi_i$ represents the phase of the $i$th oscillator, 
$M_{ij}$ the (rotational) inertia matrix, $\gamma$ the damping parameter, 
and $J_{ij}$ measures the coupling strength 
between the oscillators $i$ and $j$. 
The right-hand side describes the periodic driving with frequency $\Omega$: 
$I_i \equiv I_{i,d} + I_{i,a} \cos\Omega t$, 
where $I_{i,d}$ and $I_{i,a}$ are the amplitudes of the dc and ac components, 
respectively, of the driving on the $i$the oscillator. 
There are two cases depending on the detailed form of the inertia matrix. 
The simple case that $M_{ij}=M\delta_{ij}$ describes the 
system of coupled oscillators, each of which possesses inertia $M$
and suffers from dissipation of strength $\gamma M$ under driving $I_i$.
On the other hand, 
with $\phi_i$ denoting the phase of the superconducting order parameter
at site $i$, Eq.~(\ref{eom}) describes the dynamics of 
the array of resistively and capacitively shunted junctions (RCSJs),
where the combined direct and alternating current $I_i$
is fed into the grain at site $i$~\cite{sub,RCSJ}.
In this case, $M_{ij}$ corresponds to the capacitance matrix and
assumes the form $M_{ij} = C\Delta_{ij}$
with the junction capacitance $C$ and the lattice Laplacian
$\Delta_{ij} \equiv z\delta_{ij}- \delta_{ij'}$,
where $j'$ represents the neighboring sites of $j$
and $z$ the number of such neighbors. 
The damping parameter $\gamma$ is inversely proportional
to the shunt resistance of the junction.
For simplicity, we henceforth concentrate on the case $M_{ij}=M\delta_{ij}$
since the generalization to the case of $M_{ij} = C\Delta_{ij}$ 
is straightforward.

Equation~(\ref{eom}) may be written in the form of
Hamilton's canonical equations:
$\dot{\phi}_i = \partial H /\partial p_i$ and 
$\dot{p}_i = -\partial H /\partial \phi_i$ 
with the Caldirola-Kanai Hamiltonian~\cite{Caldirola,Denman} 
\be \label{CK}
H_{CK} = \frac{1}{2M} \sum_i (p_i +Q_i )^2 e^{-\gamma t}
      - \sum_{i<j} J_{ij} \cos (\phi_i -\phi_j ) e^{\gamma t},
\ee
where $(\phi_i , p_i)$ are conjugate variables 
and $Q_i$ is the ``gauge charge'' given by $\dot{Q}_i e^{-\gamma t}= I_i$ or
\be \label{gc}
Q_i = \frac{I_{i,d}}{\gamma} e^{\gamma t} + \frac{I_{i,a}}{\gamma^2 +\Omega^2}
        e^{\gamma t} (\gamma \cos\Omega t +\Omega\sin \Omega t)
        + Q_i^0
\ee
with arbitrary constant $Q_i^0$.
In this sense the classical mechanical system with dissipation, 
governed by the equations of motion (\ref{eom}),
can be described by the Hamiltonian in Eq.~(\ref{CK})~\cite{comm}.
We then introduce new variables $(\theta_i, \tilde{p}_i )$ according to
$p_i = \partial\Phi /\partial\phi_i$ and 
$\theta_i = \partial\Phi/\partial\tilde{p}_i$
with the generating function
$\Phi (\{\tilde{p}_i\}, \{\phi_i\}) 
\equiv \sum_i \tilde{p}_i ( \phi_i + a_i )$,
where $M\dot{a}_i \equiv -e^{-\gamma t}Q_i$. 
Under this canonical transformation, Eq.~(\ref{CK}) yields, apart from a
constant term, the new Hamiltonian
\begin{eqnarray} 
\tilde{H}_{CK} &\equiv& H_{CK} + \frac{\partial\Phi}{\partial t} \nonumber \\
    &=& \frac{1}{2M} \sum_i \tilde{p}_i^2 e^{-\gamma t} 
    - \sum_{i<j} J_{ij}e^{\gamma t}\cos (\theta_i -\theta_j -a_{ij}), \label{new}
\end{eqnarray}
where $\tilde{p}_i \,(=p_i)$ is conjugate to $\theta_i$. 
In view of the RCSJ array, where a uniform current is usually
fed into the sites along one edge and extracted from those along the opposite edge,
we consider the case that some oscillators are driven by 
$I=I_d +I_a \cos\Omega t$ 
and some others by $-I$. 
Then the bond angle $a_{ij} \equiv a_i -a_j$, depending on $(i, j)$, 
either vanishes or becomes
\be \label{a} 
a_{ij} =  \mp \frac{1}{M} 
 \left[ \frac{I_d}{\gamma} t +\frac{I_a}{\gamma^2 +\Omega^2}
         \left(\frac{\gamma}{\Omega} \sin\Omega t -\cos\Omega t \right)\right]
\ee
apart from an arbitrary constant.

Note that the energy of the system is given by 
$e^{-\gamma t}\tilde{H}_{CK}$~\cite{Denman}; 
this also corresponds to the effective Hamiltonian 
describing the statistical mechanics of the system.
To see this, we for the moment consider the system at finite temperatures
and generalize the equations of motion~(\ref{eom}) appropriately:
\be \label{eom:noise}
\sum_{j=1}^N [M_{ij} \ddot{\phi}_j + \gamma M_{ij} \dot{\phi}_j 
    + J_{ij} \sin (\phi_i -\phi_j ) ]
  = I_i + \eta_i,
\ee
where $\eta_i$ is the random (thermal) noise acting on the $i$th oscillator.
In the system of oscillators with $M_{ij}=M\delta_{ij}$,
the noise is characterized by the zero mean and the correlations
$$
\langle \eta_i (t) \eta_j (t')\rangle = 2\gamma Mk_BT \delta_{ij}\delta (t{-}t')
$$
at temperature $T$. 
In the RCSJ array with $M_{ij} = C\Delta_{ij}$,
$\eta_i$ is given by the sum of the noise currents from 
neighboring sites:
$\eta_i = \sum_j \eta_{ij}\delta_{ij'}$
with 
$$
\langle \eta_{ij}(t) \eta_{kl}(t')\rangle 
 = 2\gamma Ck_BT (\delta_{ik}\delta_{jl}-\delta_{il}\delta_{jk})\delta (t{-}t').
$$

Motivated by the canonical transformation in the absence of noise, 
we write Eq.~(\ref{eom:noise}) in the form
\begin{eqnarray} \label{Langevin}
\dot{\theta}_i &=& M^{-1} p_i e^{-\gamma t}  \nonumber \\
\dot{p}_i &=& - \sum_j J_{ij} e^{\gamma t} \sin (\theta_i -\theta_j -a_{ij})
                + \eta_i e^{\gamma t},
\end{eqnarray}
where $\theta_i \equiv \phi_i + a_i$.
The set of Langevin equations (\ref{Langevin}) may be transformed into
the Fokker-Planck equation~\cite{FP}
\begin{eqnarray} 
\frac{\partial P}{\partial t} 
 = &-&\sum_i \frac{p_i}{M}e^{-\gamma t} \frac{\partial P}{\partial \theta_i}
   + \sum_{ij} J_{ij}e^{\gamma t} \sin (\theta_i -\theta_j -a_{ij})
                       \frac{\partial P}{\partial p_i} \nonumber \\
   &+& \gamma Mk_B T \sum_{i}e^{2\gamma t} \frac{\partial^2 P}{\partial p_i^2}, 
  \label{FP}
\end{eqnarray}
which describes the time evolution of the probability distribution
$P(\{\theta_i\}, \{p_i\}, t)$ of phases and momenta at time $t$. 
Equation (\ref{FP}) yields the stationary solution valid
in the limit $t\rightarrow \infty$:
$$
P(\{\theta_i\}, \{p_i\}) \propto e^{-H_{eff}/k_B T},
$$
where the {\em effective} Hamiltonian is given by
\be \label{eff}
 H_{eff} = \frac{1}{2\tilde{M}} \sum_i p_i^2 
    - \sum_{i<j} J_{ij} \cos (\theta_i -\theta_j -a_{ij})
\ee
with $\tilde{M} \equiv Me^{2\gamma t}$.
It is thus concluded that the stationary distribution has the form
of a Gibbs measure, with the effective Hamiltonian 
indeed corresponding to the energy of the system
($H_{eff} =e^{-\gamma t}\tilde{H}_{CK}$). 

The first term in Eq.~(\ref{eff}) becomes vanishingly small 
in the stationary state $(t\rightarrow \infty)$;
it is further obvious that
the kinetic energy in the above classical system 
decouples from the interaction energy. 
We thus obtain the classical $XY$ Hamiltonian~\cite{footnote}
\be \label{ham}
H_{XY} = - J \sum_{\langle i,j\rangle} 
              \cos (\theta_i -\theta_j -a_{ij}),
\ee
where the nearest-neighbor coupling $(J_{ij}=J\delta_{ij'})$
has been assumed for convenience.
At zero temperature, which is our concern, the system described by
the Hamiltonian (\ref{ham}) is equivalent to a tight-binding particle
(of charge $e$),
with $2k_B T/J$ taking the role of the energy eigenvalue~\cite{Shih}. 
The Hamiltonian describing such a tight-binding system has
the position representation
\be \label{tb}
\langle i|H|j\rangle = e^{-ia_{ij}}\delta_{ij'} ,
\ee
where $|i\rangle$ is the position eigenket
and $a_{ij}$ may be viewed as the line integral of the 
appropriate gauge potential ${\bf a}$:
$a_{ij} = (e/\hbar c) \int_i^j {\bf a}\cdot d{\bf l}$.
Equation~(\ref{a}) shows that $a_{ij}$, defined modulo $2\pi$, 
is periodic in time with period $\tau = 2\pi m/\Omega$ only if 
\be \label{period}
I_d = \frac{s}{m} \gamma M \Omega 
\ee
with $m$ and $s$ integers.

When Eq.~(\ref{period}) is satisfied, 
the Hamiltonian (\ref{tb}) as well as the gauge field $a_{ij}$ 
has the periodicity $\tau$ and the Floquet theorem is applicable
to the corresponding Schr\"odinger equation for the wave function
$\Psi_i \equiv \langle i|\Psi\rangle$,
giving the relation~\cite{Floquet}
\be
\Psi_i (t{+}\tau) = e^{-i\tilde{E}\tau} \Psi_i(t)
\ee
with the quasienergy $\tilde{E}$.
This imposes that, apart from the dynamical contribution $\tilde{E}\tau$, 
the corresponding change in the phase of the 
wave function $\Psi_i$ should be an integer multiple of $2\pi$: 
$\theta_i (t{+}\tau)-\theta_i (t) = 2n_i \pi - \tilde{E}\tau$,
where $n_i$ is an integer depending on the form of $a_{ij}$, i.e.,
of the driving $I_i$
and $\tilde{E}$ has been assumed to be real (see below). 
We thus have the average change rate of the phase
\be \label{theta}
\langle \dot{\theta}_i \rangle 
   \equiv \frac{1}{\tau} \int_0^{\tau} dt\,\dot{\theta}_i
   = \frac{n_i}{m}\Omega - \tilde{E}
\ee
or in terms of the original phase $\phi_i$,
\be \label{phi}
\langle \dot{\phi}_i \rangle 
   = \langle \dot{\theta}_i \rangle + \frac{I_{i,d}}{\gamma M}
   = \frac{\tilde{n}_i}{m}\Omega - \tilde{E},
\ee
where $\tilde{n}_i = n_i \pm s$ or $n_i$ for driven ($I_{i,d}=\pm I_d$) 
or undriven ($I_{i,d}=0$) oscillators, respectively. 
Accordingly, the average change rate of the phase difference
or the average (dc) phase velocity,
which usually gives the appropriate physical quantity, 
e.g., the  voltage in the case of an RCSJ array, 
indeed displays subharmonic mode locking:
\be \label{lock}
\langle V_{ij}\rangle 
   \equiv \langle \dot{\phi}_i \rangle - \langle \dot{\phi}_j \rangle 
   = \frac{n}{m}\Omega  
\ee
with $n\equiv \tilde{n}_i -\tilde{n}_j$.
Note that for given configuration of driving the integer $n$ 
in Eq.~(\ref{lock}) 
is determined by the winding number $n_i$, manifesting the topological
nature of the mode locking.

The subharmonic mode locking given by Eq.~(\ref{lock}) can persist even
for the (dc) driving slightly off the condition in Eq.~(\ref{period}).
To see this, we take
\be 
\frac{I_d}{\gamma M\Omega} = \frac{s}{m} + \epsilon
\ee
for small $\epsilon$,
which leads to the bond angle either zero or
$a_{ij} = a_{ij}^0 \mp \epsilon\Omega t$
with $a_{ij}^0$ representing the periodic part for $\epsilon =0$.
In this case the Hamiltonian of the system as well as $a_{ij}$
is in general not periodic. 
However, in terms of the shifted phase
$\chi_i \equiv \theta_i +\epsilon_i \Omega t$,
where $\epsilon_i = \pm\epsilon$ or $0$ for driven or undriven 
oscillators, 
the Hamiltonian in Eq.~(\ref{ham}) reads
\be \label{ham2}
H_{XY} = - J \sum_{\langle i,j\rangle} 
              \cos (\chi_i -\chi_j -a_{ij}^0 ),
\ee
where periodicity has been restored.  Accordingly, the argument leading to
Eq.~(\ref{theta}) is applicable, giving
$\langle \dot{\chi}_i \rangle = (n_i /m)\Omega - \tilde{E}$.
This in turn leads to
\begin{eqnarray}
\langle \dot{\phi}_i \rangle 
   &=& \langle \dot{\theta}_i \rangle + \frac{I_{i,d}}{\gamma M}
   = \langle \dot{\chi}_i \rangle -\epsilon_i \Omega + \frac{I_{i,d}}{\gamma M}
   \nonumber \\
   &=& \frac{\tilde{n}_i}{m}\Omega - \tilde{E},
\end{eqnarray}
which is precisely Eq.~(\ref{phi}). 
It is thus obvious that Eq.~(\ref{lock}) remains unchanged:
Given subharmonic quantization can persist in the finite
interval of the dc driving strength $\delta I_d \propto \epsilon$,
thus generating the step structure in the appropriate response characteristics.
Since rational numbers form a dense set, this also indicates 
that there can exist multiple quantization states with different values of
$n/m$ in Eq.~(\ref{lock}) for given driving strength $I_d$. 
However, some of those, corresponding to complex values of the quasienergy
$\tilde{E}$ with nonzero imaginary parts, 
are unstable~\cite{Minorsky,Floquet}, and
only stable states (with real values of $\tilde{E}$) 
among those determine the step
structure in the actual response characteristics.  
Although such information as stability cannot be obtained by the
topological argument, 
it is certainly plausible to have two or more states stable 
in some intervals of the driving strength.
Such multistability in general gives rise to hysteresis behavior 
in the response characteristics, which has widely been observed in 
the oscillator systems with inertia terms~\cite{sub,hys,FP,Tanaka}. 
Note also that the topological argument is in essence a zero-temperature
analysis like in Refs.~\cite{Choi,qhe}:
At finite temperatures thermal fluctuations make 
the mapping onto Eq.~(\ref{tb}) and the following argument inexact. 
Nevertheless, at sufficiently low temperatures, the phase slippage
induced by fluctuations should have an exponentially low rate, 
hardly affecting the quantization itself determined by the winding number.
On the other hand, it is expected that fluctuations tend to 
destabilize various quantization states, reducing the stability interval
and the emergence of multistability. 
This is consistent with smoothing out of the step structure~\cite{GSS}
and suppression of hysteresis~\cite{Hong},
observed in the presence of noise.

In summary, we have constructed a topological argument for
subharmonic mode locking in a driven system of coupled oscillators with inertia.
Starting from the dynamic equations of motion, we have derived the 
effective Hamiltonian for the system in the
form of a classical $XY$ model with periodic bond angles, 
which in turn has been mapped onto a tight-binding particle in a 
periodic gauge field.
It has then been shown that subharmonic quantization of 
the average (dc) phase velocity follows 
as a manifestation of topological invariance, 
revealing the topological nature of the dynamical mode locking.
Also revealed is the possibility of multistability,
providing a natural explanation of the ubiquity of hysteresis
due to the inertia. 
In view of the fact that the set of equations of motion (1) describes 
a prototype
of oscillatory system,
we believe the result of this paper to be rather general 
and applicable to a variety of oscillatory systems displaying mode locking phenomena.


MYC thanks D.J. Thouless for the hospitality during his stay at University of 
Washington.
This work was supported in part by the National Science Foundation Grant
DMR-9815932 and by the Ministry of Education of Korea through the BK21 Program.

\end{multicols}


\begin{references}

\bibitem{modelock} 
 M.H. Jensen, P. Bak, and T. Bohr, Phys. Rev. Lett. {\bf 50}, 1637 (1983);
 Phys. Rev. A {\bf 30}, 1960 (1984).

\bibitem{JJ} 
For a list of references, see, e.g., {\it Proceedings of 
the 2nd CTP Workshop on Statistical Physics: KT Transition and 
Superconducting Arrays}, edited by D. Kim, J.S. Chung, and M.Y. Choi 
(Min-Eum Sa, Seoul, 1993); 
{\it Macroscopic Quantum Phenomena and Coherence in Superconducting Networks},
 edited by C. Giovannella and M. Tinkham (World Scientific, Singapore, 1995); 
{\it Proceedings of the ICTP workshop on Josephson Junction Arrays}, 
edited by H.A. Cerdeira and S.R. Shenoy, Physica B {\bf 222}, 253 (1996).

\bibitem{SS}
 S. Shapiro, Phys. Rev. Lett. {\bf 11}, 83 (1963).  
 For the subharmonic steps in the junction with finite capacitance, see 
 V.N. Belykh, N.F. Pedersen, and O.H. Sorensen, Phys. Rev. B {\bf 16}, 4860 (1977).

\bibitem{GSS}
 S.P. Benz, M. Rzchowski, M. Tinkham, and C.J. Lobb, Phys. Rev. Lett. {\bf 64},
 693 (1990);
 K.H. Lee, D. Stroud, and J.S. Chung, {\it ibid.} {\bf 64}, 962 (1990).

\bibitem{FGSS}
 In an applied magnetic field, corresponding to $k/q$ flux quanta per 
 plaquette (with $k$ and $q$ relatively prime integers), the system in
 Ref.~\cite{GSS} also displays fractional steps with the (fixed) denominator
 $q$.  Note the difference from the subharmonic steps, the denominators of 
 which are arbitrary integers.
 To aviod possible confusion, we confine our investigation here to the 
 system in the absence of a magnetic field.

\bibitem{sub}
 P.H.E. Tiesinga, T.J. Hagenaars, J.E. van Himbergen, and J.V. Jos\'e,
 J. Phys.: Condens. Matter {\bf 9}, 1813 (1997);
 M. Lee and M.Y. Choi, unpublished.

\bibitem{Arnold}
 V.I. Arnol'd, Izv. Akad. Nauk Ser. Mat. {\bf 25}, 21 (1961) 
 [Transl. Amer. Math. Soc. {\bf 46}, 213 (1965)];
 Uspekhi Mat. Nauk {\bf 38}, 189 (1983) [Russian Math. Surveys {\bf 38}, 215 (1983)].

\bibitem{Minorsky}
 N. Minorsky, {\it Nonlinear Oscillations} (Van Nostrand, Princeton, 1962);
 C. Hayashi, {\it Nonlinear Oscillations in Physical Systems} (McGraw-Hill,
 New York, 1964).

\bibitem{hys} 
See also 
H. Hong, M.Y. Choi, J. Yi, and K.-S. Soh, Phys. Rev. E {\bf 59}, 353 (1999).

\bibitem{Choi}
 M.Y. Choi, Phys. Rev. B {\bf 46}, 564 (1992); {\bf 50}, 13875 (1994).

\bibitem{Thouless}
D.J. Thouless, {\it Topological Quantum Numbers in Nonrelativistic Physics}
 (World Scientific, Singapore, 1998).

\bibitem{qhe}
D.J. Thouless, M. Kohmoto, M.P. Nightingale, and M. den Nijs, 
Phys. Rev. Lett. {\bf 49}, 405 (1982); Q. Niu, D.J. Thouless, and Y.S. Wu,
Phys. Rev. B {\bf 31}, 3372 (1985).

\bibitem{RCSJ}
M.-S. Choi, M.Y. Choi, and S.-I. Lee, Europhys. Lett. {\bf 43}, 439 (1998).

\bibitem{Caldirola}
 H. Bateman, Phys. Rev. {\bf 38}, 815 (1931); 
 P.Caldirola, Nuovo Cimento {\bf 18}, 393 (1941);
 E. Kanai, Prog. Theor. Phys. {\bf 3}, 440 (1948).

\bibitem{Denman}
 H.H. Denman and L.H. Buch, J. Math. Phys. {\bf 14}, 326 (1973).  See also
 M.Y. Choi, K.-C. Lee, and D.I. Choi, Phys. Lett. {\bf 89}A, 1 (1983);
 D.H. Kobe, G. Reali, and S. Sienlutyez, Am J. Phys. {\bf 54}, 997 (1986).

\bibitem{comm}
 There exists subtlety in the quantization of Eq.~(\ref{CK}),
giving rise to controversy as to the quantum mechanical formulation. 
See, e.g., J.M. Cervero and J. Villarroel, J. Phys. A {\bf 17}, 2963 (1984);
C.I. Um, K.H. Yeon, and W.H. Kahng, {\it ibid.} {\bf 20}, 611 (1987); 
S. Srivastava, Vishwamittar, and I.S. Minhas, J. Math. Phys. {\bf 32}, 1510 (1991).
Here our system is classical and no such complication arises. 

\bibitem{FP}
H. Risken, {\it The Fokker-Planck Equation: Methods of Solution and Applications}
 (Springer-Verlag, Berlin, 1989).

\bibitem{footnote} 
Alternatively, one may regard Eq.~(\ref{eff}) as a quantum
Hamiltonian, which at zero temperature maps onto the effective action 
of the form in Eq.~(\ref{ham}) with the coupling given by
$\sqrt{\tilde{M}J}$.  Since the stationary state 
corresponds to the strong coupling limit 
$(\sqrt{\tilde{M}J}\rightarrow \infty)$,
there do not exist quantum fluctuations 
and the classical limit is automatically achieved.  Note also that
the extra (imaginary-time) dimension plays a trivial role in this limit.

\bibitem{Shih}
 W.Y. Shih and D. Stroud, Phys. Rev. B {\bf 28}, 6575 (1983);
 M.Y. Choi, in {\it Progress in Statistical Mechanics}, edited by C.-K. Hu
 (World Scientific, Singapore, 1988), p. 385 and references therein.

\bibitem{Floquet}
 See, e.g., E.T. Whittaker and G.N. Watson, {\it A Course of Modern Analysis}
 (Cambridge Univ. Press, Cambridge, 1952);
 C.M. Bender and S.A. Orszag, {\it Advanced Mathematical Methods for
 Scientists and Engineers} (McGraw-Hill, New York, 1978).

\bibitem{Tanaka}
 H. Tanaka, A.J. Lichtenberg, and S. Oishi, Phys. Rev. Lett. {\bf 78}, 2104 
 (1997).

\bibitem{Hong}
 H. Hong, M.Y. Choi, B.-G. Yoon, K. Park, and K.-S. Soh, J. Phys. A {\bf 32}, 
 L9 (1999).


\end{references}
\end{document}